\documentclass[english,12pt]{article}
\usepackage[cp1251]{inputenc}
\usepackage{babel}
\usepackage{amsfonts, amsmath}
\usepackage{texnames}
\usepackage[T1]{fontenc}
\usepackage{graphicx}
\newcommand{\be}{\begin{equation}} \newcommand{\ee}{\end{equation}}

\newcommand{\gsim}{\mathrel{\hbox{\rlap{\lower.55ex\hbox{$\sim$}} \kern-.3em \raise.4ex \hbox{$>$}}}}

\begin{document}

\begin{center}
{\bf The Quantum Field Theory Boundaries Applicability and  Black
Holes Thermodynamics}\\
 \vspace{5mm} Alexander
Shalyt-Margolin \footnote{E-mail: a.shalyt@mail.ru;
alexm@hep.by}\\ \vspace{5mm} \textit{Research Institute for
Nuclear Problems,Belarusian State University, 11 Bobruiskaya str.,
Minsk 220040, Belarus},
\end{center}

\begin{abstract}
This paper presents a study of the applicability boundary of the
well-known quantum field theory. Based on the results of black
hole thermodynamics, it is shown that this boundary may be lying
at a level of the energy scales much lower than the Planck.  The
direct inferences from these results are given, specifically for
estimation of a cosmological term within the scope of the quantum
field theory.
\end{abstract}
PACS: 11.10.-z,11.15.Ha,12.38.Bx
\\
\noindent Keywords: quantum field theory, black holes
thermodynamics, equivalence principle
 \rm\normalsize \vspace{0.5cm}

\section{Introduction}

The local quantum field theory (QFT) is understood as a canonical
Quantum Field Theory in flat space-time
\cite{Ryder}--\cite{Wein-2}. But in what follows it is
demonstrated that a flat geometry of space-time in the processes
of high energy physics is not ensured from the start, being based
on validity of the Einstein's Strong Equivalence Principle (EP).
However, this principle has its applicability boundaries. The
Planck scales present a natural (upper) bound for applicability
because at these scales a natural geometry of space-time,
determined locally by the particular metric
$g_{\mu\nu}(\overline{x})$, disappears due to high fluctuations of
this metric and is replaced by space-time (or quantum) foam.
\\ In \cite{Shalyt-NPCS19},\cite{Shalyt-NPCS20} the author
suggested a hypothesis that actually the real applicability
boundary of EP lies in a domain of the energies $E$ considerably
lower than the Planck energies. The principal objective of this
paper is to demonstrate that the hypothesis is true for some,
quite naturally arising, assumptions. Proceeding from the
afore-said, this condition sets the applicability boundaries for
the canonical QFT. Besides, direct inferences from the obtained
result are considered. Hereinafter, EP is understood as a Strong
Equivalence Principle.

\section{The QFT Applicability Boundaries and Equivalence Principle}

The canonical quantum field theory (QFT) \cite{Ryder}--
\cite{Wein-2}  is a local theory considered in continuous
space-time with  a plane geometry, i.e with the Minkowskian metric
$\eta_{\mu\nu}(\overline{x})$. In   reality, any interaction
introduces some disturbances, introducing an additional local
(little) curvature into the initially flat Minkowskian space
$\mathcal{M}$. Then the metric $\eta_{\mu\nu}(\overline{x})$  is
replaced by the metric
$\eta_{\mu\nu}(\overline{x})+o_{\mu\nu}(\overline{x})$, where the
increment $o_{\mu\nu}(\overline{x})$ is small. But, when it is
assumed that EP is valid, the increment $o_{\mu\nu}(\overline{x})$
in the local theory has no important role and, in a fairly small
neighborhood of the point $\overline{x}$ in virtue EP.
\\The Einstein Equivalence Principle (EP) is a basic principle not
only in the General Relativity (GR) \cite{Einst1}--\cite{Wein-1},
but also in the fundamental physics as a whole. In the standard
formulation it is as follows: (\cite{Wein-1},p.68):
 \\<<{\it at every space-time point in an arbitrary
  gravitational field it is possible to choose
{\bf a locally inertial coordinate system} such that, within  a
sufficiently small region  of the point in question, the laws of
nature take the same form as in unaccelerated Cartesian coordinate
systems in the  absence of gravitation}>>.
 \\Then in (\cite{Wein-1},p.68) <<...There is also a question,
 how small is  {\bf ''sufficiently small''}. Roughly speaking, we
 mean that  the  region must be small enough so that gravitational
 field in sensible constant  throughout  it...>>.
\\However, the statement {\bf ''sufficiently small''}
is associated with another problem. Indeed, let $\overline{x}$ be
a certain point of the space-time manifold $\mathcal{M}$ (i.e.
$\overline{x}\in\mathcal{M}$) with the geometry given by the
metric $g_{\mu\nu}(\overline{x})$. Next, in accordance with EP,
there is some {\bf sufficiently small} region $\mathcal{V}$ of the
point $\overline{x}$ so that, within $\mathcal{V}$ it is supposed
that space-time has a flat geometry with the Minkowskian metric
$\eta_{\mu\nu}(\overline{x})$.
\\In essence, {\bf sufficiently small} $\mathcal{V}$  means that the
region $\mathcal{V}^{'}$, for which
$\overline{x}\in\mathcal{V}^{'}\subset\mathcal{V}$, satisfies this
condition as well. In this way we can construct the sequence
\begin{eqnarray}\label{Equiv-2}
...\subset\mathcal{V}^{''}\subset\mathcal{V}^{'}\subset\mathcal{V}.
\end{eqnarray}
The problem arises, is there any lower limit for the sequence in formula
(\ref{Equiv-2})?
\\The answer is positive. Currently, there is no doubt that
at very high energies (on the order of Planck energies $E\approx
E_p$), i.e. on Planck scales, $l\approx l_p$ quantum fluctuations
of any metric $g_{\mu\nu}(\overline{x})$ are so high that in this
case the geometry determined by $g_{\mu\nu}(\overline{x})$ is
replaced by the ''geometry'' following from {\bf space-time foam}
that is defined by great quantum fluctuations of
$g_{\mu\nu}(\overline{x})$,i.e. by the characteristic dimensions
of the quantum-gravitational region (for example,
\cite{Wheel}--\cite{Scard1}). The above-mentioned geometry is
drastically differing from the locally smooth geometry of
continuous space-time and EP in it is no longer valid
\cite{Gar1}--\cite{Garay2}.
\\From this it follows that the region $\mathcal{V}_{\overline{r},\overline{t}}$
with the characteristic spatial dimension $\overline{r}\approx
l_p$ (and hence with the temporal dimension $\overline{t}\approx
t_p$) is the lower (approximate) limit for the sequence in
(\ref{Equiv-2}).
\\It is difficult to find the exact lower limit for the sequence in formula
(\ref{Equiv-2})--it seems to be dependent on the processes under
study.  Specifically, when the involved particles are considered
to be point, their dimensions may be neglected in a definition of
the EP applicability limit. When the characteristic spatial
dimension of a particle is $\mathbf{r}$, the lower limit of the
sequence from formula (\ref{Equiv-2}) seems to be given by the
region $\mathcal{V}_{\mathbf{r'}}$ containing the above-mentioned
particle with the characteristic dimensions
$\mathbf{r'}>\mathbf{r}$, i.e. the space EP applicability limit
should always be greater than dimensions of the particles
considered in this region. By the present time, it is known that
spatial dimensions of gauge bosons, quarks, and leptons within the
limiting accuracy of the conducted measurements $<10^{-18}m$.
Because of this, the condition $\mathbf{r'}\geq 10^{-18}m$ must be
fulfilled. In addition, the radius of interaction of particles
$\mathbf{r}_{int}$ must be taken into account in quantum theory.
And this fact also imposes a restriction on considering concrete
processes in quantum theory. However, the interactions radii of
all known processes lie in the energy scales $E\ll E_p.$
\\ At the present time there is a series of the results
demonstrating that EP may be violated at the energies $E$
considerably lower than $E_p$(for example, the quantum phenomenon
of neutrino oscillations in a gravitational background
\cite{gasperini} \cite{ahluwalia} \cite{mureika}
 and others \cite{Flamb},\cite{Xiao},\cite{Dai}).
\\As QFT is a local theory applicable {\it only} to
space-time with a flat geometry determined by the Minkowskian metric
$\eta_{\mu\nu}(\overline{x})$,the applicability boundary
 EP may be considered the applicability boundary of QFT as well.
\\{\bf Main Hypothesis}
\\It is assumed that in the general case EP, and consequently, QFT
is valid for the locally smooth space-time {\it only} if all the
energies $E$ of the particles are satisfied  the necessary
condition
\begin{eqnarray}\label{Equiv-3}
 E\ll E_p,
\end{eqnarray}
In the following section this hypothesis is proven in the assumption that
space-time foam  consists of micro black  holes ({\bf mbh})
with the event horizon radius $r\approx l_p$ and mass $m\approx
m_p$.
\\
\\{\bf Remark 2.1}
\\ Why in canonical QFT it is so important never forget
about the fact that space-time has a flat geometry, or the same
possesses the Minkowskian metric $\eta_{\mu\nu}(\overline{x})$?
Simply, in the contrary case we should refuse from some fruitful
methods and from the results obtained by these methods in
canonical QFT, in particular from Wick rotation \cite{Wein-2}. In
fact, in this case the time variable is replaced by $t\mapsto
it\doteq t_E$, and the Minkowskian metric
$\eta_{\mu\nu}(\overline{x})$ is replaced by the four-dimensional
Euclidean metric
\begin{eqnarray}\label{W-1}
ds^{2}=dt^{2}_E+dx^{2}+dy^{2}+dz^{2}.
\end{eqnarray}
Clearly, such replacement is possible only in the case when from
the start space-time (locally) has a flat geometry, i. e.
possesses the Minkowskian metric $\eta_{\mu\nu}(\overline{x})$.
This is another argument supporting the key role of the EP
applicability boundary. Otherwise, when we go beyond this
boundary, Wick rotation becomes invalid. Naturally, some other
methods of canonical QFT will lose their force too.

\section{The Strong Equivalence Principle, Black Holes, and QFT}

It is supposed that a large (i. e, classical)
four-dimensional Schwarzschild black hole is existent with the metric
\begin{equation}\label{bh-1}
ds^{2}=\left( 1-\frac{2MG}{r}\right) dt^{2}-\left(
1-\frac{2MG}{r}\right) ^{-1}dr^{2}-r^{2}d\Omega ^{2},
\end{equation}
where $M$ is  the mass of this black hole, and the Schwarzschild
horizon radius $r_{BH}$ is defined by
\begin{equation}
r_{BH}=2MG.  \label{bh-2}
\end{equation}
As shown in \cite{Singl-1},\cite{Singl-2}, EP is violated for an
observer distant from the black hole event horizon. Considering
our objective, it seems expedient to give in brief the main
results from \cite{Singl-1},\cite{Singl-2}.
\\ In view of the Unruh effect, an accelerating observer
 does detect thermal radiation (so-called Unruh radiation)
 with the Unruh temperature given by \cite{unruh}
\begin{eqnarray}\label{Un-1}
 T_{Unruh} = \frac{\hbar a}{2 \pi},
\end{eqnarray}
where $a=|{\bf a}|$ is a corresponding  acceleration.
\\When an observer is at the   {\it fixed} distance,
$r>r_{BH}$, from  a Schwarzschild black hole of mass $M$ and event
horizon radius $r_{BH}=2GM$, then, due to the existence of Hawking
radiation \cite{hawking}, the observer will measure radiation with
thermal spectrum and a temperature given by formula \cite{Singl-1}
\begin{eqnarray}\label{Un-2}
T_{H,r}=\frac{\hbar }{8 \pi G M \sqrt{1- \frac{r_{BH}}{r}}}~,
\end{eqnarray}
where $r>r_{BH}$.
\\  In the foregoing formulae and in what follows, we use the normalization
$c=k_B=1$.
\\Besides, in \cite{Singl-1} it is shown that an observer,
positioned at the fixed distance $r>r_{BH}$ from the
above-mentioned black holes and measuring Hawking temperature with
the value $T_{H,r}$, experiences the local acceleration
\begin{eqnarray}\label{Un-3}
a_{BH,r} =  \frac{1}{\sqrt{1- \frac{r_{BH}}{r}}}
\left(\frac{r_{BH}}{2r^2} \right) ~.
\end{eqnarray}
Another observer in the Einstein elevator, moving with
acceleration through Minkowskian space-time, will measure the same
acceleration toward the floor of the elevator, thermal radiation
with  the Unruh temperature given by formula (\ref{Un-1}). As
shown in \cite{Singl-1}, $a_{BH}$ is coincident with the quantity
$a$ from the formula in (\ref{Un-1}). Then substituting the
acceleration $a=a_{BH}$ from formula (\ref{Un-3}) into formula
(\ref{Un-1}), we can obtain a formula for $T_{Unruh,r}$ in this
case \cite{Singl-2}:
\begin{eqnarray}\label{Un-4}
T_{Unruh,r} = \frac{\hbar}{2 \pi \sqrt{1- \frac{r_{BH}}{r}}}\left(
\frac{r_{BH}}{2r^2}\right) ~.
\end{eqnarray}
If EP is valid, the quantities $T_{Unruh,r}$ from formula
(\ref{Un-4}) and $T_{H,r}$ in (\ref{Un-2}) should be coincident
for $r>r_{BH}$ to a high degree of accuracy. However, we see that
this is not true. In \cite{Singl-2}, e.g. for $r=4GM=2r_{BH}$, we
have $T_{H,r}=4T_{Unruh,r}$.
\\So, far from the event horizon, EP is not the case.
Moreover, violation of EP is the greater the farther it is from
the black hole event horizon. Indeed, for an observer at the
distance $r>r_{BH}$ we can write
$r=\mathcal{\alpha}(r)GM=\frac{1}{2}\mathcal{\alpha}(r)r_{BH},\mathcal{\alpha}(r)>2$.
Then
\begin{eqnarray}\label{Un-4.1}
T_{Unruh,r} = \frac{\hbar}{2\pi \mathcal{\alpha}^{2}(r)G M
\sqrt{1-\frac{2}{\alpha(r)}}}.
\end{eqnarray}
In this way $T_{H}/T_{Unruh,r}=\mathcal{\alpha}^{2}(r)/4$. And the
ratio is the greater, the higher $\mathcal{\alpha}(r)$,i.e. the
farther from horizon the observer is. Next, for compactness, we
denote $T_{Unruh,r}$ in terms of $T_{U,r}$. Of course, in this
case we bear in mind only an observer at a sufficiently great but
finite distance from a black hole, i.e. only when a gravitational
field is thought significant and must be taken into consideration.
So, in the general case $r_{BH}<r\ll\infty$, whereas in the case
of a distant observer we have
\begin{eqnarray}\label{Un-5.1}
r_{BH}\ll r\ll\infty.
\end{eqnarray}
Obviously, this case of violated EP is not directly associated
with the {\bf Main Hypothesis} concerning the boundaries of EP
validity (formula (\ref{Equiv-3})) from the previous section,
because in \cite{Singl-1},\cite{Singl-2}  consideration is given
to a large black hole with the event horizon radius  $r_{BH}$ much
greater than Planck length $r_{BH}\gg l_p$ at sufficiently low
energies.
\\Really, the resulting distribution of the particles emitted
by a black hole has the form (last formula on p.122 in
\cite{Mukhanov})
\begin{eqnarray}\label{Un-6}
n_E=\Gamma_{gb}[exp(\frac{E}{T_{H}})-1]^{-1},
\end{eqnarray}
where $n_E$ is the number of particles with the energy $E$ and
$\Gamma_{gb}<1$ is the so-called {\it greybody factor}. As the
black hole mass $M$ is large, the temperature $T_{H}$ is low, and
then from the last formula it follows that arbitrary large values
of $n_E$ will be given only by particles with the energies $E$
close to a small value of $T_{H}$.
\\The principal result from the remarkable papers
\cite{Singl-1},\cite{Singl-2} may be summarized as follows:
\\{\bf Comment 3.1}
\\In any point of space-time that is in a field of a large classical Schwarzschild black hole,
and in the cases when this field must be taken into consideration,
it is impossible to remove this field in the vicinity of the point
even locally, i.e. to consider space-time as flat.
\\
\\{\bf Comment 3.2}
\\It is important to refine some formulations from \cite{Singl-1},\cite{Singl-2}.
Specifically, if $r\rightarrow r_{BH}$, then $T_{H,r}\rightarrow
\infty,T_{Unruh,r}\rightarrow\infty$ in formulae (\ref{Un-2}) and
(\ref{Un-4}), respectively.  Note that for $r\rightarrow r_{BH}$
these temperatures become infinite
$T_{H,r}=\infty,T_{Unruh,r}=\infty$. Based on this fact, in
 \cite{Singl-1},\cite{Singl-2} it is inferred ''that the
equivalence principle is restored on the horizon''. But this
statement is not correct. Restoration of EP is not following from
the fact that the above temperatures take infinite values. We can
only state that temperature on the BH horizon and in its vicinity
cannot be the parameter detecting a deviation from EP. In the
opposite case one can arrive at violation: on the black hole event
horizon, where a gravitational field is very large in value, EP
holds, whereas far from the  event horizon, where a gravitational
field is much weaker, this principle is violated.
\\
\\Let us return to high energy physics and to the subject of the previous section.
One of the preferable models for space-time foam is the model
based on the assumption that its unit cells are {\bf mbh}, with
radius and mass on the order of the Planck (for example,
\cite{Scard1},\cite{Scard2}, \cite{Scard3}. Of great importance
for {\bf mbh} are the quantum-gravitational effects and the
corresponding quantum corrections of black hole thermodynamics at
Planck scale (for example, \cite{Nou}).
\\ Then, in line with formula (20) in \cite{Nou},
we have minimal values for radius and mass   of a black hole
\begin{equation}\label{min}
r_{min}=\sqrt{\frac{e}{2}}\alpha^{'} l_p,\quad m_{min}\doteq
m_0=\frac{\alpha^{'}\sqrt{e}}{2\sqrt{2}}m_p,
\end{equation}
where the number $\alpha^{'}$ is on the order of 1, and in
\cite{Nou} we take the normalization $\hbar=c=k_B=1$ in which
$l_p=m_{p}^{-1}=T_{p}^{-1}=\sqrt{G}.$
\\From (\ref{min}) it directly follows that the formula
for the event horizon radius $r=2MG$, valid for large classical
black holes, will be valid in the case when we include the
quantum-gravitational effects for {\bf mbh} because
$r_{min}=2m_0G$. Such a black hole of a minimal size is associated
with a maximal temperature (formula (24) in \cite{Nou}):
\begin{equation}\label{tempmax}
T_{H}^{max}=\frac{T_{p}}{2\pi \sqrt{2}\alpha^{'}}.
\end{equation}
A black hole satisfying the formulae
(\ref{min}),(\ref{tempmax}) is termed as  {\it minimal} (or {\it Planck}).
\\ Without loss of generality,
it is assumed that for event horizon radii and masses of {\bf mbh}
the following is valid:
\begin{equation}\label{min.2}
r_{mbh}\approx r_{min},m_{mbh}\approx m_0,
\end{equation}
i.e. {\bf mbh} are {\it Planck} black holes.
\\ For the energies $E$ somewhat lower than the Planck energies
(i.e., $E\ll E_p$) involved in the condition (\ref{Equiv-3}), a
semiclassical approximation is valid. This means that, on
substitution of {\bf mbh} with the mass $m_{mbh}$ for a large
(classical) black hole with the mass $M$, in the case under study
the results, substantiated when an observer uses the standard
Unruh-Dewitt detector in radiation measurement for coupled to a
massless scalar field \cite{BD},\cite{akhmedov}, are valid with
the corresponding quantum corrections \cite{Keifer}.
\\ Let us revert to the formulae from \cite{Singl-1},\cite{Singl-2}.
In particular, to formula (\ref{Un-3}) for the real
acceleration measured by an observer who is positioned at the fixed
distance $r\gg r_{BH}$ in the Schwarzschild space-time, given
by (formula (14) in \cite{Singl-1}, formula (9.170) in
\cite{carroll})
\begin{equation}\label{acc-sch}
a_{BH,r}=a_S = \frac{{\sqrt{\nabla _\mu V \nabla ^\mu V}}}{V} =
\frac{MG}{r^2 \sqrt{1-2MG/r}}=\frac{MG}{r^2V},
\end{equation}
where it is supposed that  a static observer at the radius $r$ moves
along orbits of the time-like Killing vector
$\textit{K}=\partial_{t}$ and
$V=\sqrt{-\textit{K}_{\mu}\textit{K}^{\mu}}=\sqrt{1-2MG/r}$ is the
red-shift factor for the Schwarzschild space-time (p.413 in
\cite{carroll}).
\\ How changes formula (\ref{acc-sch}) on going to {\bf
mbh}? It is clear that the condition $r\gg r_{BH}$ is replaced by
the condition $r\gg r_{mbh}$ (corresponding to the condition $E\ll
E_p$ and semiclassical approximation),$M$ is replaced by
$m_{mbh}$,the red-shift factor $V$ should be replaced by $V_q$,
where $V_{q}$ is the quantum deformation of $V$ with regard to
quantum corrections in the field {\bf mbh}. Then, for {\bf mbh},
formula (\ref{acc-sch}) is of the form
\begin{equation}\label{acc-sch.q}
a_{S,q} =\frac{m_{mbh}G}{r^2V_q},
\end{equation}
where $a_{S,q}$  is the real acceleration with regard to the
quantum corrections measured by a distant observer in the field
{\bf mbh}.
\\ Clearly, formula (\ref{Un-2}) for $T_{H,r}$, due to formulae for
 the red-shift factor  $V$, in the general case may be given as
\begin{eqnarray}\label{Un-2.new}
T_{H,r}=\frac{\hbar }{8 \pi GM V}~.
\end{eqnarray}
Then its quantum analog, i.e.
the corresponding formula for temperature in the field {\bf mbh}, for $r\gg r_{mbh}$ is as follows:
\begin{eqnarray}\label{Un-2.new.q}
T_{H,r,q}=\frac{\hbar}{8\pi Gm_{mbh}V_q}.
\end{eqnarray}
In virtue of formula (\ref{acc-sch}), formula
(\ref{Un-4}) takes the form
\begin{eqnarray}\label{Un-4.new1}
T_{U,r} = \frac{\hbar}{2 \pi \sqrt{1- \frac{r_{BH}}{r}}}\left(
\frac{r_{BH}}{2r^2}\right)=\frac{\hbar}{2 \pi V}\left(
\frac{r_{BH}}{2r^2}\right).
\end{eqnarray}
For $r>r_{BH}$ and due to formula (\ref{Un-4.1}), we have
\begin{eqnarray}\label{Un-4.1.new1}
T_{U,r}=\frac{\hbar}{2\pi \mathcal{\alpha}^{2}(r)G M
\sqrt{1-\frac{2}{\alpha(r)}}}= \frac{\hbar}{2\pi
\mathcal{\alpha}^{2}(r)G M V}.
\end{eqnarray}
As indicated above, for $r\gg r_{BH}$ we have $\alpha(r)\gg 1$.
\\ What are the changes on going to {\bf mbh}?
\\ Considering the case $r\gg r_{mbh}$ and semiclassical
picture, we again come to $\alpha(r)\gg 1$, whereas formula
(\ref{Un-4.1.new1}) is replaced by formula
\begin{eqnarray}\label{Un-4.1.new1,q}
T_{U,r,q} = \frac{\hbar}{2\pi \mathcal{\alpha}^{2}(r)G
m_{mbh}V_q}.
\end{eqnarray}
Proceeding from the above, we have \begin{eqnarray}\label{BH-1}
\frac{T_{H,r,q}}{T_{U,r,q}}=\frac{T_{H,r}}{T_{U,r}}=\frac{\mathcal{\alpha}^{2}(r)}{4}
\end{eqnarray}
Formula (\ref{BH-1}) points to the fact that, within the scope of
a semiclassical approximation, relations of a black hole
temperature to the Unruh temperature for a distant observer in the
case of a large (classical) black hole and {\bf mbh} are
coincident because these quantities are dependent on the same
factors:
\\ first, on $1/MV$ and, second, on $1/m_{mbh}V_q$.
\\ Note that in this consideration there
s no need to have an explicit formula for $V_q$ as this quantity
is not involved in the key expression (\ref{BH-1}). Specifically,
to derive an explicit expression for $V_q$, we can use the results
from \cite{Sol} on quantum deformation of the Schwarzschild
solution due to spherically symmetric quantum fluctuations of the
metric and the matter fields. In this case the Schwarzschild
singularity at $r=0$ is shifted to the finite radius
$r_{min}\approx r_{mbh}\propto l_p$, where the scalar curvature is
finite. In this way the results from \cite{Sol} correlate well
with the results from \cite{Nou}.
\\Quantum corrections at Planck scales were obtained in \cite{Nou}
proceeding from validity of the Generalized Uncertainty Principle
(GUP)\cite{GUPg1}--\cite{Tawf}. But the results presented in this
work are independent of this aspect. Actually, during studies of
black hole thermodynamics at Planck scales with the use of other
methods \cite{Corr},\cite{LQG} (differing from those in
\cite{Nou}),in particular, Loop Quantum Gravity (LQG)\cite{LQG},
the obtained results were similar to \cite{Nou}. Because of this,
for {\bf mbh} with all the thermodynamic characteristics
(mass,radius,temperature,...) on the order of the corresponding
Planck quantities, all the calculations in this section are valid.
As noted above, far from horizon of {\bf mbh}, i.e. at the
energies $E\ll E_p$ (\ref{Equiv-3}), the results from
\cite{Singl-1},\cite{Singl-2} remain valid in this case as well.
\\Next, similar to \cite{Scard1}, we assume that in every cell of
space-time foam a micro black hole ({\bf mbh}) with a typical
gravitational radius of $r_{min}\propto l_p$ may be present. Then,
in according with the results in \cite{Singl-1},\cite{Singl-2} and
in virtue of the formula (\ref{BH-1}) we come to violation of the
strong EP for distance $r$, satisfying the condition
\begin{eqnarray}\label{Un-5.2}
l_p\ll r\ll\infty,
\end{eqnarray}
that is equivalent to $\widetilde{E}_r\ll E_p$ for the energies
$\widetilde{E}_r$ associated with the scale of $r$.
\\In the last formula it is implicitly (and purely conditionally)
assumed that a minimum length is equal to $l_p$ and to $r_{min}$.
But, as noted above, in the general case we have $r_{min}\propto
l_p$, i.e., the order is similar to that of $l_p$. Specifically,
in \cite{Rong-1} by natural assumptions it has been demonstrated
that the minimum length may be twice and more as great as the
Planck length. It is obvious that all the above calculations and
derivations of the present work are independent of the specific
value of $r_{min}$.
\\In this way, if the quantum foam structure is determined by {\bf
mbh}, the applicability of QFT is limited to the energies
$E<\widetilde{E}_r\ll E_p$ and the formula (\ref{Equiv-3}) is the
case. This supports the {\bf Main Hypothesis} from Section 2
within the assumption concerning the quantum foam structure made
in this section.
\\{\bf Comment 3.3}
\\ In the case of {\bf mbh} {\bf Comment 3.2} is absolutely clear.
In fact, at a horizon of {\bf mbh},i.e. for $r=r_{min}$,
$T_{H,r}=T_{Unruh,r}=\infty$ similar to large black holes but ,
naturally, without any restoration of EP as the domain
$r=r_{mbh}\approx r_{min}\propto l_p$ is the region of Planck
energies or of quantum foam, where EP in its canonical formulation
becomes invalid. It is obvious that at the event horizon
$r=r_{mbh}$ of {\bf mbh} and in its vicinity a gravitational field
becomes very strong due to quantum effects and nothing could
destroy it.

\section{Some Immediate Consequences}

{\bf 4.1} Based on the above results, all the energies $E$ we can classify into 3 groups:
\\
\\a)low energies $0<E\leq\widetilde{E}\ll E_p$
-- energies, for which the Strong Equivalence Principle is valid
in virtue of formula (\ref{Equiv-3}), and hence this energy
interval sets the QFT applicability boundaries.
\\a1) Since $\widetilde{E}\ll E_p$,
it is natural to assume that $\widetilde{E}\approx 10^{-N}E_p$,
where $N\geq 2$. Obtaining of more accurate estimates for $N$ is a
separate problem;
\\
\\b)intermediate energies $\widetilde{E}<E<E_p$ -- energies,
for which the Strong Equivalence Principle and, consequently QFT,
becomes invalid but the corresponding scales are greater than the
Planck. It can be assumed that QFT in this energy range will be a
theory in a gravitational field that could not be destroyed even
locally. In the case under study it is assumed that this field is
created by {\bf mbh}. Impossibility of destroying this field even
locally is associated with large quantum corrections for the
corresponding quantities which should be taken into consideration
at these energies \cite{Nou},\cite{Corr},\cite{LQG}.
\\ Let us call the energy scale $\widetilde{E}<E< E_p$ as
{\it prequantum gravity phase};
\\
\\c)high (essentially maximal) energies $E\approx E_p$ or
$E>E_p$. This interval is the region of quantum gravity energies.
\\
\\ Next note that, as all the experimentally involved energies $E$ are low,
they satisfy condition a) or b). Specifically, for LHC, maximal
energies are $\approx 10TeV=10^{4}GeV$, that is by 15 orders of
magnitude lower than the Planck energy $\approx 10^{19}GeV$.
Moreover, the characteristic energy scales of all fundamental
interactions also satisfy condition a). Indeed, in the case of
strong interactions this scale is $\Lambda_{QCD}\sim 200MeV$; for
electroweak interactions this scale is determined by the vacuum
average of a Higgs boson and equals   $\upsilon\approx 246GeV$;
finally, the scale of the (Grand Unification Theory (GUT))
$M_{GUT}$ lies in the range of $\sim10^{14}GeV--10^{16}GeV$.
\\ It should be noted, however, that on validity of assumption a1)
the energy scale $M_{GUT}$ lies within the applicability region of
the energy group a) and hence of QFT. Provided the EP
applicability boundaries are lying at considerably lower energies,
a study of GUT necessitates a theory with (even locally)
unremovable curvature.
\\At the same time, it is clear that the requirement of the
Lorentz-invariant QFT, due to the action of Lorentz boost (or same
hyperbolic rotations) (for example formula (3) in \cite{Akhm}),
results in however high momenta and energies. But it has been
demonstrated that unlimited growth of the momenta and energies is
impossible because in this case we fall within the energy region,
where the conventional   quantum field theory
\cite{Ryder}--\cite{Wein-2} is invalid.
\\Note that at the present time there are experimental indications
that Lorentz-invariance is violated in QFT on passage to higher
energies (for example, \cite{Kost}). Besides, one should note
important recent works associated with EP applicability boundaries
and violation in nuclei and atoms at  low energies (for example
\cite{Flamb}). Proceeding from the above, the requirement for
Lorentz-invariance and EP is possible only within the scope of the
condition (\ref{Equiv-3}).
\\
\\{\bf 4.2} Proceeding from the above results,
it is inferred that the well-known QFT
\cite{Ryder}--\cite{Wein-2}, from the start, is a
ultraviolet-finite theory with the natural cutoff parameter
$l_{\widetilde{E}}\propto \hbar/\widetilde{E}.$ Note that the
quantum-gravitational parameter $r_{mbh}\propto l_p$ is beyond the
applicability limits of QFT.
\\
\\{\bf 4.3} In the present approach it is of interest to study
the problem of {\it asymptotic safety} introduced by Steven
Weinberg in \cite{Wein-UV}. We use the definition of this notion
given in (\cite{Keifer--book},p.67): ''A theory is said to be
asymptotically safe if all essential coupling parameters $g_j$
(these are the ones that are invariant under field redefinitions)
approach, for energies $k\rightarrow\infty$, a fixed point where
at least one of them does not vanish.'' If initially it has been
assumed that $r_{min}$ is considered within the scope of
(GUP)\cite{GUPg1}--\cite{Tawf},this definition necessitates
certain refinements. In particular, if (GUP) supposes the
existence of a maximum momentum $p_{max}$, as in \cite{Nozari} or
(Section V in \cite{Tawf}),it is clear that the condition
$k\rightarrow\infty$ can not be fulfilled. So, the condition
$k\rightarrow\infty$ should be replaced by the condition
$k\rightarrow p_{max}.$ Most often,  it is assumed that the
momentum $p_{max}$ is on the order of the Planck momentum,  i.e.,
we have $p_{max}\propto p_{pl}$. However, in the most general case
the quantity $p_{max}$ may be even trans-Planck.
\\ When $p_{max}$ is inexistent (i.e. $p_{max}=\infty$),
still by this approach of  {\it asymptotic safety} the problem
should be reformulated in accordance with the fact (shown above)
that, beginning with the energies $E,E>\widetilde{E}$, a theory
must be considered as QFT in curved space-time in a field created
by {\bf mbh}.
\\In his further works the author is planning to study this problem within the scope of this approach in greater detail.
\\
\\{\bf 4.4} It is possible to correct the estimates
obtained within the scope of the known QFT by means of the
condition (\ref{Equiv-3}). Let us consider a typical example.
\\In his well-known lectures \cite{Wein-cosm} at the Cornell University
Steven Weinberg considered an example of calculating, within the
scope of QFT, the expected value for the vacuum energy density
$<\rho>$ that is proportional to the cosmological term $\lambda$.
To this end, zero-point energies of all normal modes of some field
with the mass $m$ are summed up to the wave number cutoff
$\Lambda\gg m$ for the selected normalization $\hbar=c=1$ (formula
(3.5) in \cite{Wein-cosm}):
\begin{eqnarray}\label{Equiv-4}
<\rho>=\int_0^\Lambda\frac{4\pi
k^{2}dk}{(2\pi)^{3}}\frac{1}{2}\sqrt{k^{2}+m^{2}}\simeq
\frac{\Lambda^{4}}{16\pi^{2}}.
\end{eqnarray}
Assuming, similar to \cite{Wein-cosm}, that GR is valid at all the
energy scales up to the Planck, we have the cutoff $\Lambda\simeq
(8\pi G)^{-1/2}$ and hence (formula (3.6) in \cite{Wein-cosm})
leads to the following result:
\begin{eqnarray}\label{Equiv-5}
<\rho>\approx 2\cdot 10^{71}GeV^{4},
\end{eqnarray}
that by $10^{118}$ orders of magnitude differs from the well-known
experimental value for the vacuum energy density
\begin{eqnarray}\label{Equiv-6}
<\rho_{exp}>\preceq 10^{-29}\mathrm{g}/cm^{3}\approx
10^{-47}GeV^{4}.
\end{eqnarray}
Here $G$ is a gravitational constant.
\\It is clear that in this case the condition (\ref{Equiv-3})
is not fulfilled and this leads to such a monstrous discrepancy
with $<\rho_{exp}>$. Based on the afore-said, the following
estimate for $<\rho>$ is more correct:
\begin{eqnarray}\label{Equiv-4.0}
<\rho_{\widetilde{E}}>\doteq\int_{0}^{\Lambda_{\widetilde{E}}}\frac{4\pi
k^{2}dk}{(2\pi)^{3}}\frac{1}{2}\sqrt{k^{2}+m^{2}}\simeq
\frac{\Lambda_{\widetilde{E}}^{4}}{16\pi^{2}},
\end{eqnarray}
where $\Lambda(\widetilde{E})$--cut-off parameter of the
corresponding energy $\widetilde{E}$  from point a) in {\bf 4.1}.
\\ Of course, the main contribution into the integral
in the right side of (\ref{Equiv-4}) is made by high energies
$\widetilde{E}<E< E_p$ from point b) in {\bf 4.1},which are not
involved in formula (\ref{Equiv-4.0}). Consequently, it seems
possible that $<\rho_{\widetilde{E}}>\ll <\rho>$ and hence
$<\rho_{\widetilde{E}}>$ may be much closer to $<\rho_{exp}>$ than
$<\rho>$.
\\ In the whole, the methods of a quantum field theory may be effectively used to
 obtain the vacuum energy density. In his very interesting work
\cite{Rong-2} the author, based on the quantum hoop conjecture,
has obtained a natural cutoff for the vacuum energy of a scalar
field, also giving an estimate for $<\rho>$ much closer to
$<\rho_{exp}>$ than in  formulae(\ref{Equiv-4}),(\ref{Equiv-5}).
It is interesting to know how close are the estimates for the
vacuum energy density in \cite{Rong-2} and in formula
(\ref{Equiv-4.0}). To this end, we first need a sufficiently
accurate estimate of the quantity $\widetilde{E}$ in line with the
QFT boundaries applicability. Besides, it is important to find
whether the methods of QFT are enough to obtain $<\rho_{exp}>$ or
some additional assumptions will be required, specifically,  the
Holographic Principle applying  to the Universe
\cite{Sussk1}--\cite{shalyt-IJMPD} .

\section{Conclusion}

Thus, within the scope of natural assumptions, in this paper it is
demonstrated that the applicability boundary of the well-known QFT
is lying in the region of energies considerably lower than the
Planck energies, i.e. the canonical QFT
\cite{Ryder}--\cite{Wein-2} is an ultraviolet-finite theory.
\\In this paradigm it is important to understand the way to transform
the well-known results for ultraviolet regularization,
renormalization, and so on from QFT within the scope of the
applicability boundary $\widetilde{E}$ of QFT indicated in point
a) of the preceding section. Possibly, this boundary will be
dependent on a nature of the processes under study in high energy
physics.

\begin{center}
{\bf Conflict of Interests}
\end{center}
The author declares that there is no conflict of interests
regarding the publication of this paper.
\\
\begin{center}
{\bf Acknowledgments}
\end{center}
The author would like to express his gratitude to the Reviewers of this work for their valuable remarks and recommendations.

\end{document}